\documentclass{article}
\usepackage{spconf,amssymb,amsmath,epsfig}
\usepackage{textcomp}
\usepackage{amsfonts,amsthm,amsopn}
\usepackage{bm}
\usepackage{hyperref}
\usepackage{url}
\usepackage[capitalise]{cleveref}
\usepackage{graphicx,caption,lipsum}
\usepackage{booktabs,multirow}
\usepackage{placeins}
\usepackage{algpseudocode,algorithm}
\usepackage{color}
\usepackage[table,xcdraw]{xcolor}
\usepackage{bbm}
\usepackage{enumerate}   
\usepackage{adjustbox}
\usepackage[normalem]{ulem}
\usepackage{fixltx2e}
\usepackage[toc,page]{appendix}

\title{DEAAN: Disentangled Embedding and Adversarial Adaptation Network for Robust speaker representation learning}

\name{{Mufan Sang, Wei Xia, John H.L. Hansen}}
\address{
Center for Robust Speech Systems, The University of Texas at Dallas, TX, USA \\
{\small \tt \{mufan.sang, wei.xia, john.hansen\}@utdallas.edu}}

\newcommand{\vct}[1]{\boldsymbol{\mathbf{#1}}} % vector
\graphicspath{{./image/}}

\begin{document}

\maketitle

\ninept

\begin{abstract}
Despite speaker verification has achieved significant performance improvement with the development of deep neural networks, domain mismatch is still a challenging problem in this field.  In this study, we propose a novel framework to disentangle speaker-related and domain-specific features and apply domain adaptation on the speaker-related feature space solely. Instead of performing domain adaptation directly on the feature space where domain information is not removed, using disentanglement can efficiently boost adaptation performance. To be specific, our model's input speech from the source and target domains is first encoded into different latent feature spaces. The adversarial domain adaptation is conducted on the shared speaker-related feature space to encourage the property of domain-invariance. Further, we minimize the mutual information between speaker-related and domain-specific features for both domains to enforce the disentanglement. Experimental results on the VOiCES dataset demonstrate that our proposed framework can effectively generate more speaker-discriminative and domain-invariant speaker representations with a relative 20.3\% reduction of EER compared to the original ResNet-based system. 
\end{abstract}
%
%------------------------------------------------------------------------------------------
%------------------------------------------------------------------------------------------

\vspace{-1ex}
\section{Introduction}

The task of speaker verification (SV) is to identify the true characteristics of the speaker from audio streams and to accept or discard the identity claimed by the speaker~\cite{hansen2015speaker}. As the previous state-of-the-art method, i-vector has been far surpassed by current neural network-based deep speaker embedding systems~\cite{zhang2017end,xia2020speaker,wan2018generalized,snyder2018x,cai2018exploring}. Among them, x-vector and ResNet-based architectures are most widely used. Although current deep speaker systems trained on a large amount of labeled data can extract speaker-discriminative and robust speaker embeddings, their performance may still degrade significantly when deployed to new unseen domains~\cite{villalba2019state}. It is caused by the domain mismatch between training and unseen test data. Deep speaker embedding networks can improve the generalization ability with large-scale in-domain datasets. In practice, however, it might be very costly or even impossible to collect sufficient data annotations.

Many extrinsic factors can lead to performance degradation, such as channel mismatch, language, environmental noise, and room reverberation. To alleviate this problem, numerous robust speaker verification systems have been proposed. One approach used in~\cite{snyder2018x,villalba2019state} is to improve the robustness of speaker verification systems with data augmentation by adding noise, music, and reverberant speech. However, data augmentation has its constraint that it can only provide specific acoustic variations.

The goal of domain adaptation (DA) is to reduce mismatch across domains. It has been performed on the back-end level, input-level, and embedding-level. For back-end level, PLDA can be adapted with unlabeled in-domain data on its mean and covariance matrix. For input-level, models usually can be adapted by training with enhanced~\cite{shi2020robust} or domain-translated~\cite{nidadavolu2019low} input features. For adaptation at embedding-level, it often targets at minimizing certain distances between source and target domains to align them in the same embedding space, such as cosine distance~\cite{sang2020open}, mean squared error (MSE)~\cite{cai2020within}, and maximum mean discrepancy (MMD)~\cite{lin2020multi}. However, this method usually requires parallel or artificial simulated data, which cannot generalize well to real-world scenarios.

Recently, adversarial learning becomes a popular approach to adapt speaker representations. For instance,~\cite{wang2018unsupervised,xia2019cross} explored domain adversarial learning to learn domain-invariant and speaker-discriminative representations. Similar ideas are also applied in~\cite{zhou2019training,meng2019adversarial,wang2020few}, but these systems need more extra noise labels or environment labels, which are hard to obtain in real-world scenarios. However, these DA methods have a common limitation that they implement the adaptation typically on the feature space, which encodes speaker-related and speaker-unrelated factors collectively. Therefore, the adaptation on feature space is unavoidably interfered with the speaker-independent factors, even leading to negative transfer~\cite{pan2009survey}. Although~\cite{kwon2020intra} introduced their study on speaker representation disentangling with Voxceleb datasets, the cross-dataset adaptation with disentangling is still an open question to investigate. 

\begin{figure*}[th]
  \centering
    %\hspace{-15mm}
    \vspace{-2mm}
    \includegraphics[width=18cm,height=7.0cm]{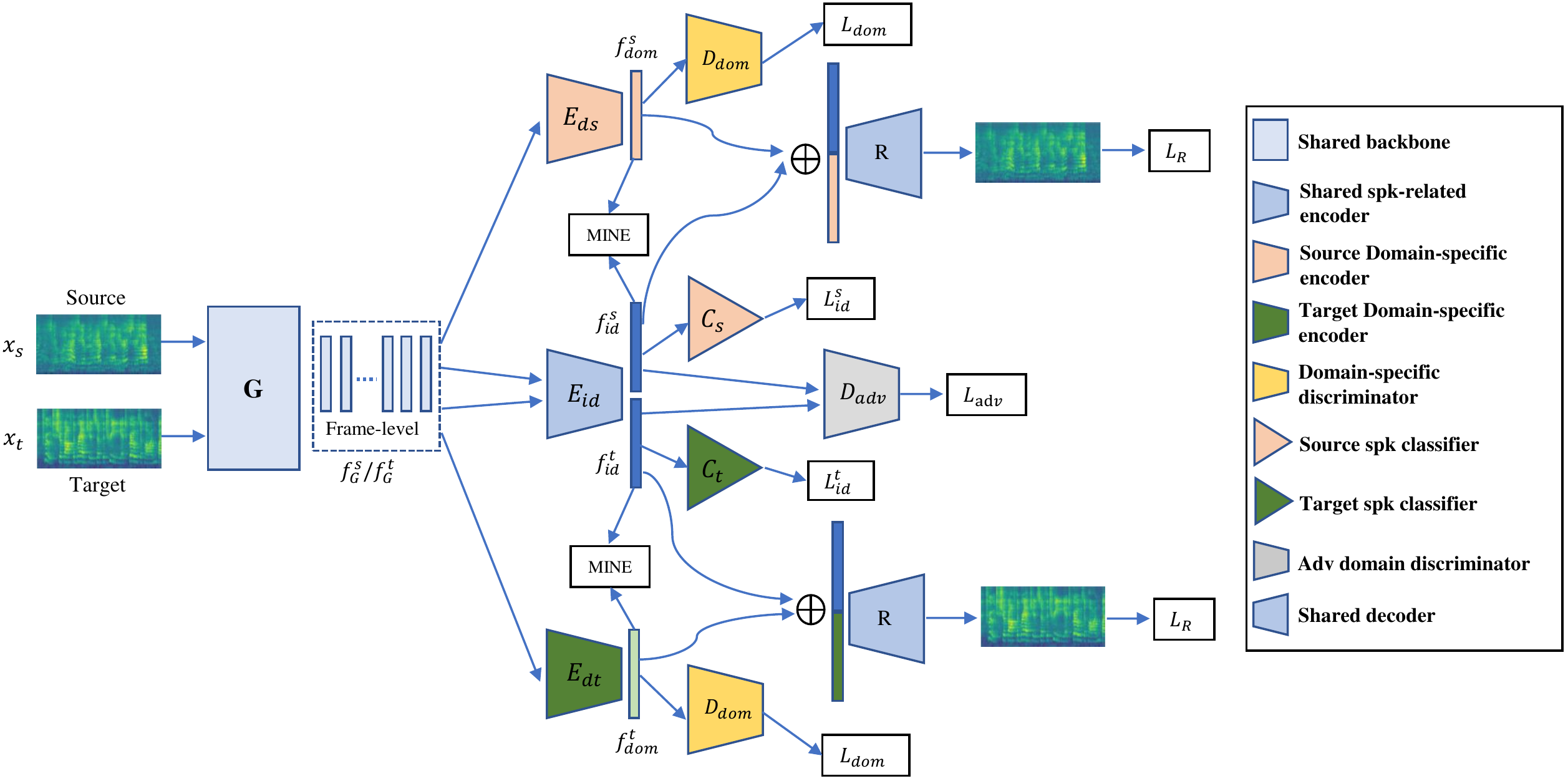}
    \vspace{-3.1mm}
  \caption{Overview of the proposed Disentangled Embedding and Adversarial Adaptation Network} 
    %\vspace{-3mm}
\end{figure*}

In this work, we propose a novel framework Disentangled Embedding and Adversarial Adaptation Network (DEAAN), aiming for the cross-domain speaker verification problem by disentangling speaker-related and the undesired domain-specific features. Unlike previous methods treating adaptation and disentangling separately, we incorporate both of them into our framework and train them jointly. In this way, domain adaptation can be operated more effectively on the speaker-related space without the interference of domain-specific features. Accordingly, the proposed framework aims to encode the source and target inputs into three latent feature spaces by the corresponding shared speaker-related encoder and two domain-specific encoders, as shown in Figure 1. The shared speaker-related space contains speaker-related features across domains. The source and target domain-specific feature spaces capture domain-specific features only. Meanwhile, a domain adversarial loss is applied to encourage the encoder to produce domain-invariant features. To guarantee the disentanglement, we propose to minimize the mutual information between disentangled speaker-related features and domain-specific features in both domains. Our proposed approach makes disentangling and adaptation mutually benefit each other. The disentangled feature enables the adaptation to concentrate on speaker-related features exclusively, and adaptation correspondingly enhances the disentangling with a more domain-invariant encoder. We evaluate our proposed framework on a far-field dataset Voices Obscured in Complex Environmental Settings (VOiCES)~\cite{richey2018voices}. Experimental results show that our framework can significantly improve speaker verification performance using the same backbone structure as the ResNet-based system utilized in this study.

In Sec. 2, we describe the details of our proposed DEAAN framework. Data description and detailed experiment settings, results and discussions are reported in Sec. 3 and 4. Finally, we conclude our work in Sec. 5.

\label{sec:intro}

\section{Disentangled Embedding and Adversarial Adaptation Network}
In this section, we introduce our proposed approach of DEAAN with the goal of obtaining robust disentangled speaker embeddings which contain only speaker-related information. To make the speaker embedding robust to domain variations, we collectively consider the adversarial domain adaptation and representation disentanglement to decouple domain-invariant speaker representations from input speech signals. Therefore, our framework incorporates disentanglement and adversarial domain adaptation to extract speaker-related and domain-invariant speaker representations.

The entire model architecture of DEAAN is illustrated in Figure 1. It can be mainly concluded as two modules: 1) disentanglement,  and 2) domain adaptation. To be specific, disentangling module consists of following components: a shared speaker  frame-level feature extractor (referred as backbone $G$), a speaker-related encoder $E_{id}$, two domain-specific encoders $E_{ds}$ and $E_{dt}$, a domain discriminator $D_{dom}$, two speaker identity classifiers $C_{s}$ for source domain, and $C_{t}$ for target domain, and a decoder $R$. The adaptation module consists of an adversarial domain discriminator $D_{adv}$. 

For the backbone, we use a ResNet34 architecture. As shown in Figure 1, the extractor $G$ extracts frame-level latent features  $\vct{f}_{G}$ from input features. The three encoders $E_{id}$, $E_{ds}$, $E_{dt}$ aim to disentangle the frame-level latent features into speaker-related features  and domain-specific features with speaker identity labels. The decoder $R$ is responsible for reconstructing the original input feature ($\vct{x}_s$ or $\vct{x}_t$) from a fusion of speaker-related and domain-specific feature. In order to promote the disentanglement, we minimize the mutual information between disentangled features to decrease their dependency. The adversarial domain discriminator $D_{adv}$ is trained to align the distributions of speaker-related features $\vct{f}_{id}$ across source and target domains. With the adversarial training strategy, our proposed adaptation network can effectively extract robust and domain-invariant speaker features. 

\vspace{-0.5ex}
\subsection{Speaker embedding network backbone}
In this work, we use the ResNet34 with self-attentive pooling (SAP)~\cite{cai2018exploring} as our original speaker recognition system. This speaker embedding network can encode variable length of utterance into a fix-length speaker representation. This system is trained using cross-entropy (CE) as the speaker identity loss. Meanwhile, we use the ResNet34 only as our backbone $G$ for DEAAN to extract frame-level latent features.

\vspace{-0.5ex}
\subsection{Embedding disentanglement}
Although the original speaker recognition system above can extract discriminative speaker embeddings, there are still some speaker-unrelated information entangled in the feature space. To decouple the speaker-related features from embeddings, we enforce the shared encoder $E_{id}$ to encode speaker-related information from speaker embeddings across domains, and encourage the two separated domain-specific encoders to capture their speaker-unrelated domain-specific features. As depicted in Figure 1, given a source input and a target input, the speaker-related feature for source $\vct{f}_{id}^s$ and target $\vct{f}_{id}^t$ going through speaker classifiers $C_{s}$ and $C_{t}$ are expected to preserve the strong speaker discriminative power with the speaker identity loss,

\begin{equation}
L_{\mathrm{id}}=\mathbb{E}\left[-\sum_{i=1}^{I} y_{i}^{s} \log C_{{s}}\left(\vct{f}_{i d}^{s}\right)_{i}\right] + \mathbb{E}\left[-\sum_{j=1}^{J} y_{j}^{t} \log C_{{t}}\left(\vct{f}_{id}^{t}\right)_{j}\right]
\end{equation}
Where $y_{i}^{s}$ and $y_{j}^{t}$ denotes the class of the $i$-th sample in source domain and the $j$-th sample in target domain.

To ensure domain-specific features are disentangled by the two encoders $E_{ds}$ and $E_{dt}$ properly, a shared domain discriminator $D_{dom}$ is applied to discriminate the domain-specific feature $\vct{f}_{dom}^s$ or $\vct{f}_{dom}^t$ based on their domain membership (source or target). $D_{dom}$ is trained by minimizing the following binary cross-entropy (BCE) loss function,

\begin{equation}
\begin{aligned}
L_{\mathrm{dom}} &=\mathbb{E}\left[-\log D_{d o m}\left(\vct{f}_{dom}^s \right)\right] \\
&+\mathbb{E}\left[-\log D_{d o m}\left(1-\vct{f}_{dom}^t\right)\right]
\end{aligned}
\end{equation}

\noindent
Disentangled embeddings are expected to preserve information of the original input, thus the decoder $R$ is required to reconstruct the original inputs $\vct{x}_{s}$ and $\vct{x}_{t}$ in source and target domains from the fusion of ($\vct{f}_{id}^s$, $\vct{f}_{dom}^s$) and ($\vct{f}_{id}^t$, $\vct{f}_{dom}^t$), respectively. This can be achieved by minimizing the distance between reconstructed input and original input. We utilize $L_{2}$ distance as the distance metric, the reconstruction loss function is formulated as,
\begin{equation}
\begin{aligned}
L_{\mathrm{R}} &=\mathbb{E}\left[\left\|R\left(\vct{f}_{i d}^{s}, \vct{f}_{d o m}^{s}\right)-\vct{x}_{s}\right\|_{2}^{2}\right] \\
&+\mathbb{E}\left[\left\|R\left(\vct{f}_{i d}^{t}, \vct{f}_{d o m}^{t}\right)-\vct{x}_{t}\right\|_{2}^{2}\right]
\end{aligned}
\end{equation}
In this way, disentangled embeddings are enforced to retain more information of original inputs.   

\subsection{Mutual information minimization}

To achieve better disentanglement performance, we minimize the mutual information between speaker-related and domain-specific features on source and target domains. Theoretically, mutual information is represented as the Kullback-Leibler (KL) divergence between the joint distribution ${\mathbb{P}_{X Z}}$ (refered as $\mathbb{J}$) and the product of marginals ${\mathbb{P}_{X} \otimes {\mathbb{P}}_{Z}}$ (referred as $\mathbb{M}$) of two random variables $X$ and $Z$. In~\cite{belghazi2018mutual}, a mutual information neural estimation (MINE) was proposed to estimate the mutual information for high dimensional continuous variables based on a deep learning architecture. With the Donsker-Varadhan representation~\cite{donsker1983asymptotic}, mutual information can be estimated as,
\begin{equation}
\begin{aligned}
\mathcal{I}_{\Theta}(X ; Z) &=D_{\mathrm{KL}}(\mathbb{J}|| \mathbb{M}) \\
&=\sup _{\theta \in \Theta} \mathbb{E}_\mathbb{J}\left[T_{\theta}\right]-\log \left(\mathbb{E}_\mathbb{M}\left[e^{T_{\theta}}\right]\right)\label{con:DV}
\end{aligned}
\end{equation}
% \begin{equation}
% I \widehat{(X ; Z})_{n}=\sup _{\theta \in \Theta} \mathbb{E}_{\mathbb{P}_{X Z}^{(n)}}\left[T_{\theta}\right]-\log \left(\mathbb{E}_{\mathbb{P}_{X}^{(n)} \otimes \widehat{\mathbb{P}}_{Z}^{(n)}}\left[e^{T_{\theta}}\right]\right)
% \end{equation}
Where $T_{\theta}$ : $\mathcal{X} \times \mathcal{Z} \rightarrow \mathbb{R}$
denotes a neural network parameterized by $\theta$ in the equation. Mutual information between continuous random variables can be estimated by maximizing the above equation.
Considering we primarily concern minimizing mutual information instead of its detailed value, we propose to apply an alternative MINE which is more stable based on the Jensen-Shannon divergence (JSD)~\cite{nowozin2016f}. The Equation \ref{con:DV} can be alternatively formulated as,
\begin{equation}
\mathcal{I}_{\Theta}(X ; Z)=\mathbb{E}_\mathbb{J}\left[-\operatorname{sp}\left(-T_{\theta}\right)\right]-\mathbb{E}_\mathbb{M}\left[\operatorname{sp}\left(T_{\theta}\right)\right]\label{con:JSD}
\end{equation}
Where $\mathrm{sp}(z)$=log(1+$e^z$) represents a softplus function.

The information overlap between ($\vct{f}_{id}^s$, $\vct{f}_{dom}^s$) as well as ($\vct{f}_{id}^t$, $\vct{f}_{dom}^t$) can be minimized by minimizing their mutual information.  
Thus, we achieve embedding disentanglement by playing a min-max game. First, MINE (refered as $M$) is trained to estimate mutual information between pairs of features ($\vct{f}_{id}^s$, $\vct{f}_{dom}^s$), and ($\vct{f}_{id}^t$, $\vct{f}_{dom}^t$) by maximizing $\mathcal{I}_{\Theta}$ in Equation \ref{con:JSD}. Meanwhile, encoders are trained to produce speaker-related and domain-specific features with minimum mutual information. Thus, we define the corresponding loss as,   
\begin{equation}
L_{\mathrm{MI}}=\min _{E_{id},E_{ds},E_{dt}} \max _{M} \sum_{i=1}^{N} \mathcal{I}_{\Theta}\left(\left(\vct{f}_{i d}\right)_{i} ; \left(\vct{f}_{dom}\right)_{i}\right) 
\end{equation}
Where $\vct{f}_{i d}$ and $\vct{f}_{dom}$ represent ($\vct{f}_{id}^{s}$ or $\vct{f}_{id}^{t}$) and ($\vct{f}_{dom}^{s}$ or $\vct{f}_{dom}^{t}$). 
A gradient reversal layer (GRL)~\cite{ganin2016domain} is adopted to play the min-max game for this objective function.

\subsection{Adversarial domain alignment}
In embedding disentanglement, we apply the strategy that shares the weights for the backbone extractor and speaker-related encoder between source and target domains. 
%~\cite{liu2017unsupervised,tzeng2017adversarial} 
But this strategy cannot ensure that the speaker-related features are properly adapted and have similar distribution across domains. To encourage the characteristic of domain-invariant for disentangled speaker-related features, we exploit an adversarial domain discriminator $D_{adv}$ which aims to distinguish the domain of speaker-related features $\vct{f}_{id}^s$  and $\vct{f}_{dom}^s$. On the other hand, the encoder $E_{id}$ learns to confuse the $D_{adv}$ and generates speaker-related features which cannot be differentiated by the discriminator. The adversarial loss can be presented as,

\begin{equation}
\begin{aligned}
L_\mathrm{a d v} &=\mathbb{E}\left[ \log D_{a d v}\left(\vct{f}_{i d}^{s}\right)+ \log \left(1-D_{a d v}\left(\vct{f}_{i d}^{t}\right)\right)\right] \\
&+\mathbb{E}\left[\log D_{a d v}\left(\vct{f}_{i d}^{t}\right)+ \log \left(1-D_{a d v}\left(\vct{f}_{i d}^{s}\right)\right)\right]
\end{aligned}
\end{equation}
With this designed mechanism, the shared encoder $E_{id}$ is able to extract domain-invariant and speaker-related features.

\subsection{Overall optimization}
In summary, the overall objective function of DEAAN is a weighed sum of all loss functions above represented as,

\vspace{-0.1ex}
\begin{equation}
\begin{array}{c}
L_{\mathrm{DEAAN}}=L_{\mathrm{i d}}+\lambda_\mathrm{d o m} L_{\mathrm{d o m}}+\lambda_\mathrm{adv} L_{\mathrm{a d v}}+\lambda_\mathrm{{R}} L_{R} +\lambda_\mathrm{{M I}} L_\mathrm{{M I}}
\end{array}
\end{equation}
Where {$\lambda_\mathrm{{dom}}$}, {$\lambda_\mathrm{{adv}}$}, {$\lambda_\mathrm{{R}}$}, and {$\lambda_\mathrm{{MI}}$} are hyper-parameters to control the weight for each composition. In our experiments, they are set as: {$\lambda_\mathrm{{dom}}= 0.5$}, {$\lambda_\mathrm{{adv}}= 0.5$}, {$\lambda_\mathrm{{R}}= 0.2$}, and {$\lambda_\mathrm{{MI}}= 0.2$}.

\section{Experimental setting}

\subsection{Datasets}
\subsubsection{Source domain data}
We use the development set of Voxceleb1 and Voxceleb2 (refer to as Vox1-dev and Vox2-dev) which were collected “in the wild”~\cite{nagrani2017voxceleb,chung2018voxceleb2} as our source domain data. The former contains 148642 utterances from 1211 speakers, and the latter one comprises of around 1 million utterances from 5994 speakers. We do not use any data augmentation strategies in experiments. 
\vspace{-1mm}
\subsubsection{Target domain data}
For target domain, we use the Voices Obscured in Complex Environmental Settings (VOiCES) Challenge 2019 corpus~\cite{richey2018voices} which is challenging with speech recorded by various distance far-field microphones in different noisy rooms. The corpus consists of development set (refer to as VOiCES-dev) with 15904 utterances from 196 speakers, and evaluation set (refer to as VOiCES-eval) with 11392 utterances from more challenging recording conditions. Our proposed DEAAN is evaluated on the VOiCES-eval set with 3.6 million trials. We report the system performance with two evaluation metrics: Equal Error Rate (EER) and minimum Detection Cost Function (DCF) with {$P_{{target}}= 0.01$}.

\begin{table}[]
\caption{Speaker verification results on the VOiCES evaluation set with different adaptation methods.}
\vspace{-2mm}
\setlength{\tabcolsep}{1.3mm}{
\renewcommand\arraystretch{1}
\begin{tabular}{cccc}
\bottomrule
\textbf{System}               & \textbf{Train set}                                                        & \textbf{EER (\%)}       & \textbf{minDCF}        \\ \hline \midrule
\multirow{2}{*}{ResNet34-SAP} & \multirow{2}{*}{\begin{tabular}[c]{@{}c@{}}Vox1+Vox2\end{tabular}} & \multirow{2}{*}{7.52} & \multirow{2}{*}{0.532} \\
                              &                                                                           &                         &                        \\ \hline
ResNet34-SAP-FT               & \begin{tabular}[c]{@{}c@{}}Vox1+Vox2\\ +VOiCES-dev\end{tabular}       & 6.54                  & 0.458                  \\ \hline
Peri {~\cite{peri2020robust}}                   & \begin{tabular}[c]{@{}c@{}}Vox1+Vox2 \end{tabular}                  & 9.07                  & N/A                     \\ \hline
ADSAN+MINE {~\cite{yi2020adversarial}}             & \begin{tabular}[c]{@{}c@{}}Vox1+Vox2+Aug\\ +VOiCES-dev\end{tabular}       & 6.76                  & 0.599                  \\ \hline
Our proposed DEAAN                        & \begin{tabular}[c]{@{}c@{}}Vox1+Vox2\\ +VOiCES-dev\end{tabular}   & \textbf{5.21}                  & \textbf{0.394}                  \\ \bottomrule
\end{tabular}}
\vspace{-3mm}
\end{table}
%\vspace{-4mm}
\subsection{Feature extraction and ResNet-based models training}
For all systems, 64-dimensional log-mel filter bank energies are extracted with a frame-length of 25 ms at a 10 ms shift. We apply mean-normalization over a sliding window of up to 3 seconds, and Kaldi energy-based VAD is used to remove silence frames.

We use the ResNet34 with self-attentive pooling (SAP) method trained by the CE loss function as the original system. It is trained on the Vox1-dev and Vox2-dev without any data augmentation. For each training step, we randomly make a chunk of 300 to 800 frames of each utterance to keep samples in each minibatch sharing the same frame length.  

Based on the pre-trained ResNet34-SAP model, we fine-tune it on the VOiCES-dev with the replaced new last classification layer and all other layers (referred as ResNet34-SAP-FT). Same settings for input and loss function as the above system are adopted during fine-tuning. We evaluate the fine-tuned model on the VOiCES-eval. 
% which contains 3.6 million pairs in the trial.
\vspace{-2.5mm}
\subsection{DEAAN implementation details}
For our proposed architecture, we use the Vox1-dev and Vox2-dev as the source domain data and development set of VOiCES as the target domain data to train the DEAAN. We exploit a ResNet34 as the backbone to extract the frame-level features. The shared speaker-related encoder and two domain-specific encoders share the same structure which consists of one SAP layer and two fully-connected layers. The decoder $R$ reconstructs the original input feature through three fully-connected layers and nine transposed convolution layers~\cite{radford2015unsupervised} with batch normalization.

For training, we use the SGD optimizer with a learning rate of 1e-4 and momentum of 0.9 to optimize the adversarial domain discriminator. Adam optimizer is applied with an initial learning rate of 1e-4 decreasing by 5\% every 5 epochs to train other parts. We train the model adversarially with the procedure similar in~\cite{lee2020drit++}. Different from the ResNet34-SAP system and the fine-tuned one ResNet34-SAP-FT, we use a shorter fixed length input with 384 frames randomly cropped from each utterance to train our DEAAN. The back-end comprises of LDA for reducing the embedding dimension to 200, centering, length normalization, and a Gaussian PLDA.

\begin{figure}[t]
    \begin{minipage}[b]{0.3\linewidth}
        \centering
        \centerline{\includegraphics[height=2.8cm, width=2.8cm]{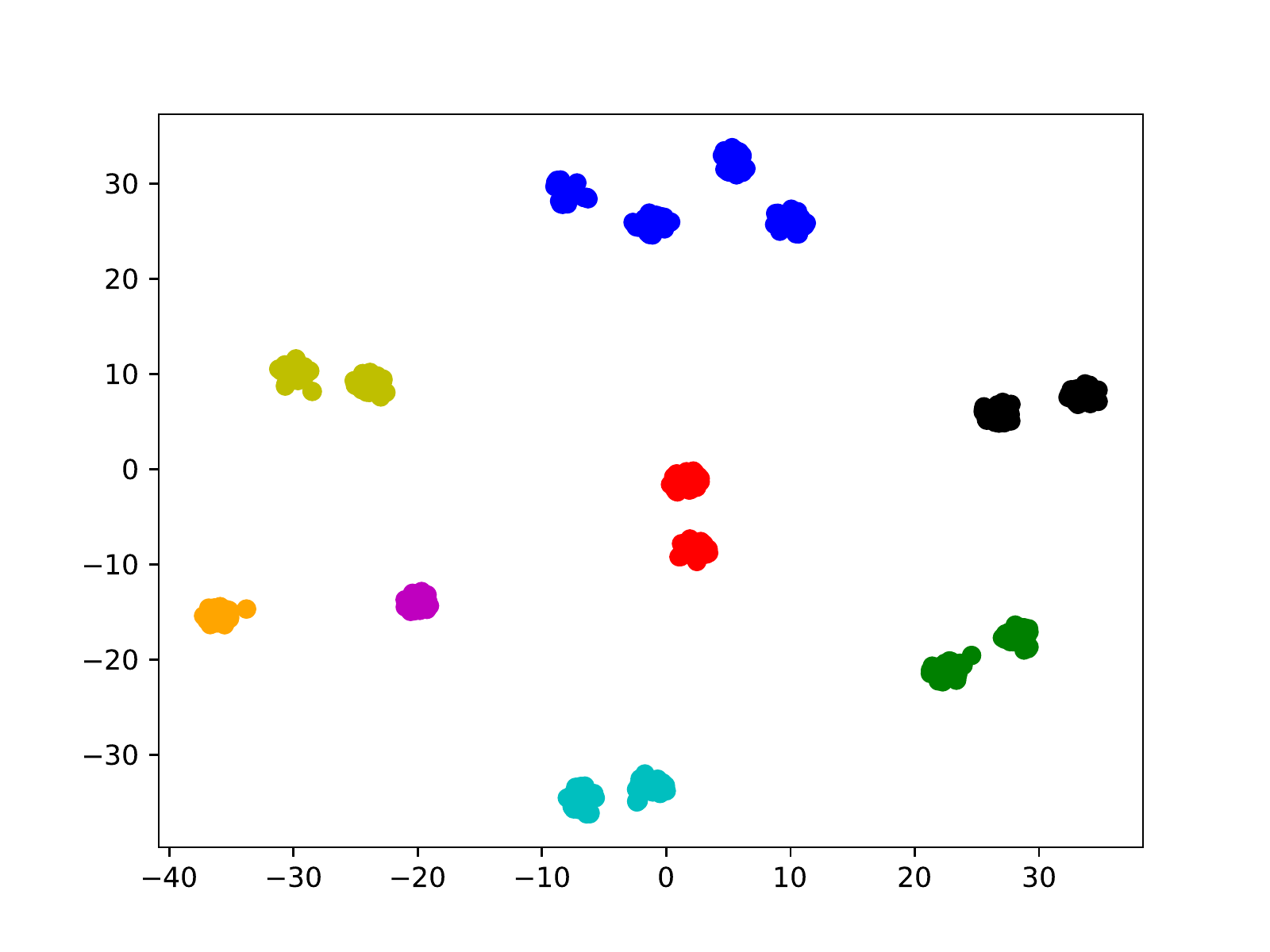}}
        \centerline{(a) ResNet34-SAP-FT }\medskip
      \end{minipage}
      \hfill
    \begin{minipage}[b]{0.3\linewidth}
        \centering
        \centerline{\includegraphics[height=2.8cm, width=2.8cm]{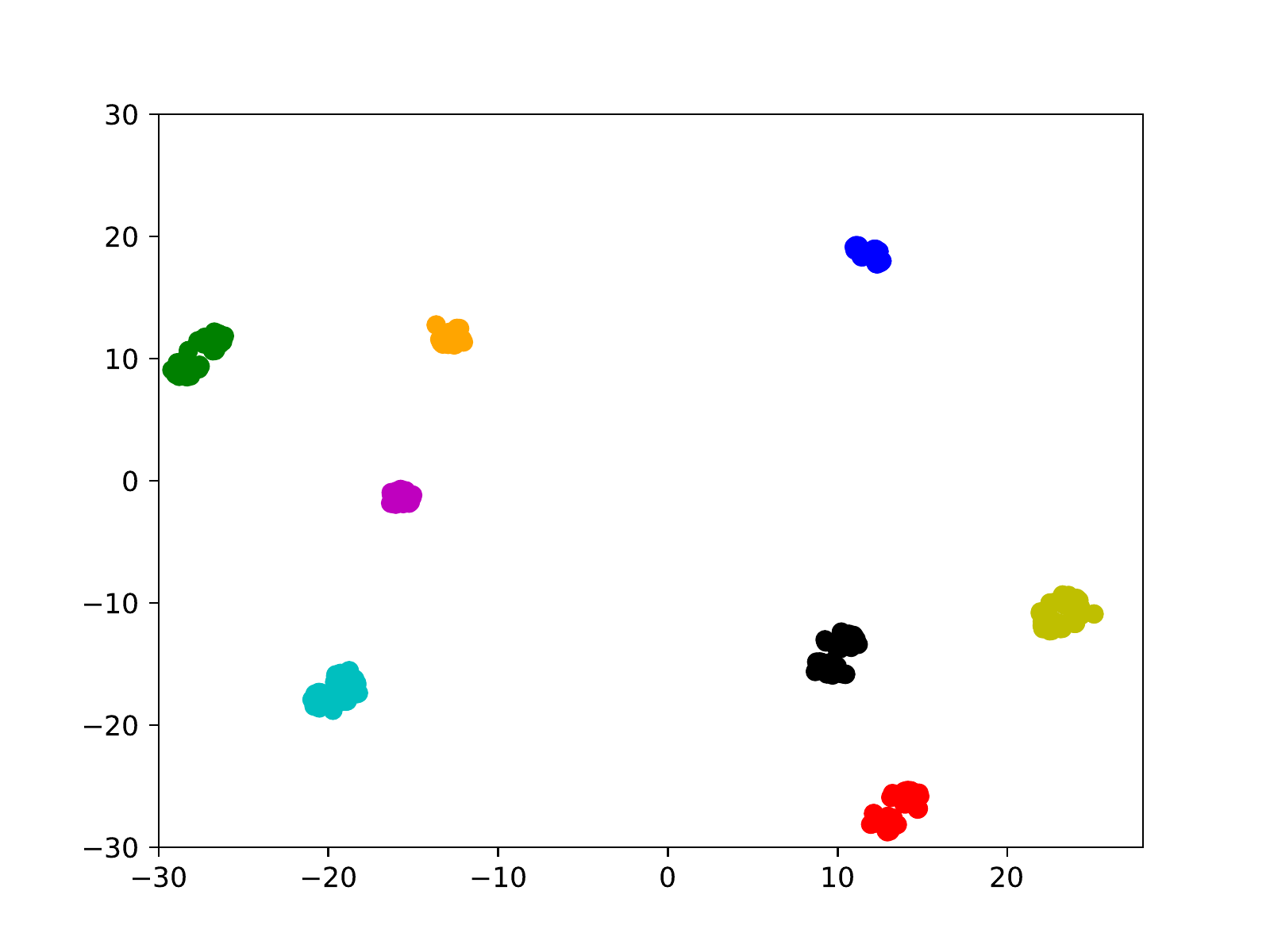}}
        \centerline{ (b)DEAAN spk-related }\medskip
      \end{minipage}
      \hfill      
    \begin{minipage}[b]{0.3\linewidth}
      \centering
      \centerline{\includegraphics[height=2.8cm, width=2.8cm]{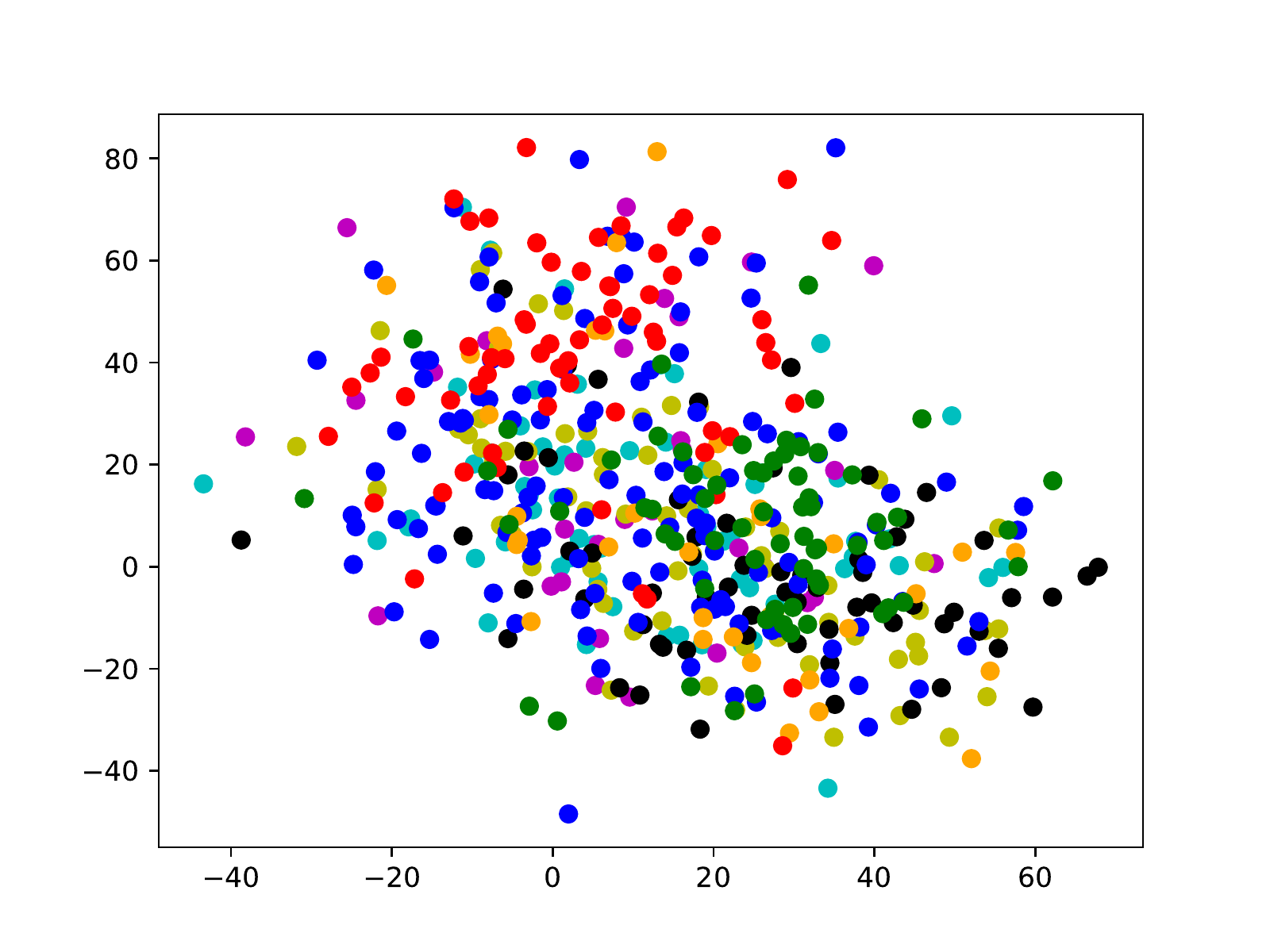}}
      \centerline{(c) Domain-specific}\medskip
    \end{minipage}
    \vspace{-3mm}
    \caption{t-SNE plots of extracted embeddings for 8 speakers: (a) Speaker embeddings extracted from the ResNet34-SAP-FT. (b) speaker-related features extracted from DEAAN. (c) domain-specific features extracted from DEAAN}
    \vspace{-4mm}
    \label{fig:tsne}
\end{figure}
%\vspace{-1ex}
\vspace{-2.0ex}
\section{Results and Discussions}
%\vspace{-5.0ex}
We investigate the performance of our DEAAN architecture and compare it with the ResNet34-SAP and the fine-tuned ResNet34-SAP-FT. 

Table 1 shows system performance on the VOiCES-eval. Methods~\cite{peri2020robust,yi2020adversarial} are recently proposed adversarial training architectures for speaker representation learning. As shown in the Table, although the proposed DEAAN is trained with shorter input (384 frames) and use the same frame-level extractor as the ResNet34-SAP and ResNet34-SAP-FT, it outperforms both of them significantly with a relative 30.7\% and 20.3\% reduction in EER, and 25.9\% and 13.9\% reduction in minDCF, respectively. With the best performance produced by DEAAN in Table 1, it meets our assumption that the incorporation of our proposed adversarial domain adaptation and embedding disentanglement can effectively improve the speaker verification performance. With the same classification loss and even shorter length of input, the results demonstrate that our model can generate more speaker-discriminative and domain-invariant speaker embeddings even in challenging environments. 

To visualize the effectiveness and disentanglement performance of the proposed model, we use the t-distributed Stochastic Neighbor Embedding (t-SNE) plots to illustrate the performance improvement in Figure 2. Speakers are randomly chosen from target domain. Figure 2 (a) is the visualization of embeddings extracted from the ResNet34-SAP-FT, Figure 2 (b) represents speaker-related features $\vct{f}_{id}$ extracted from DEAAN, and Figure 3 (c) shows the corresponding domain-specific features.
As shown in Figure 2 (a), it can be observed that the identity clusters are not compact enough (e.g., blue cluster) and even contain small separated clusters in each identity. It may can be explained that some recorded speech segments share similar or same noise in the room and far-field conditions, but the entire recording process was conducted with various channels and noise variabilities. Compared to the ResNet34-SAP-FT, the speaker-related features extracted by our proposed model are denser for same identity and the gap caused by domain or channel mismatch is diminished. Moreover, domain-specific features shown in Figure 3 (c) are distributed uniformly in the feature space and independent to speaker identities.
Therefore, the above results indicates that our proposed method can effectively generate discriminative and domain-invariant speaker-related features.
\vspace{-2.0ex}
\section{Conclusions}
\vspace{-1.0ex}
In this paper, we propose a novel speaker embedding framework that disentangles speaker-related and domain-specific features and performs domain adaptation on the speaker-related feature space without detailed mismatch information by adversarial training. By introducing a mutual information neural estimator and adversarial domain adaptation, DEAAN can effectively adapt speaker embeddings across domains without domain-related interference. Our approach incorporates representation disentanglement and adversarial domain adaptation together, which enable them to mutually benefit each other. Experimental results demonstrate that our approach can produce more speaker-discriminative and domain-invariant features to improve the speaker verification performance in challenging environments compared to the original and fine-tuned systems with using the same backbone structure.

\vfill\pagebreak

% \newpage

\bibliographystyle{IEEEbib}
\bibliography{icassp21}

% \newpage
% \onecolumngrid
% \appendix
% \section*{Supplementary Material}

\end{document}